# Measuring AI Systems Beyond Accuracy

## Violet Turri[1], Rachel Dzombak[2], Eric Heim[3], Nathan VanHoudnos[4], Jay Palat[5], Anusha Sinha[6]

Carnegie Mellon University Software Engineering Institute
vmturri@sei.cmu.edu[1], rdzombak@sei.cmu.edu[2], etheim@sei.cmu.edu[3], nmvanhoudnos@sei.cmu.edu[4],
vpalat@sei.cmu.edu[5], asinha@sei.cmu.edu[6]

**Abstract**

Current test and evaluation (T&E) methods for assessing machine learning (ML) system performance often rely on incomplete metrics. Testing is additionally often siloed from the other phases of the ML system lifecycle. Research investigating cross-domain approaches to ML T&E is needed to drive the state of the art forward and to build an Artificial Intelligence (AI) engineering discipline. This paper advocates for a robust, integrated approach to testing by outlining six key questions for guiding a holistic T&E strategy.

## Introduction

Most machine learning projects focus on "accuracy" for model evaluation. While accuracy is useful for knowing how well a model performs on a test dataset at the time of model development, there are other significant implications in assessing the utility and usability of a machine learning model. Key considerations include robustness, resilience, calibration, confidence, alignment with evolving user requirements, and fit for mission and stakeholder needs as part of an integrated system, among others. In this paper, we explore what it means to measure beyond accuracy and define critical considerations for the test and evaluation of machine learning and, more broadly, artificial intelligence systems. After defining six key considerations related to robust T&E, the AI engineering community will be better equipped to develop and implement comprehensive and applied methods for the evaluation of models as well as possible metrics for more realistic and real-world model evaluation.

## Current AI T&E Practices

Modern AI systems, many of which are built using machine learning, are a departure from static software systems that yield deterministic results. In contrast to analytical systems that follow explicit "instructions" given by a programmer and can be reduced and decomposed, AI systems are empirical, opaque, and unpredictable — they behave based on what they "learn" from data or experience (Russell & Norvig, 2021). Because AI systems often model real-world relationships, they must be adaptable to changing inputs and shifting correlations.

The characteristic differences between traditional software and AI lead to a series of open questions around the design and implementation of AI. For example, AI systems can evolve and change behavior over time. How can we ensure that AI systems are still doing what they are supposed to do? How can we certify that they are safe and reliable? Many current AI and ML methods are data intensive; continuous updates to data, while necessary, can impact architectural concerns such as prediction accuracy and latency (Ozkaya, 2020). Technical debt, driven by data dependencies, accumulates rapidly, silently, and at the system level (Sculley et al., 2015). What practical mechanisms exist to evaluate the state of a system when using large and evolving data sets? Further, exhaustive testing is currently not possible for systems that learn and adapt. How do we change approaches to test and evaluation to be more risk-, resilience-, and process- focused rather than exhaustive?

A common practice for engineering AI systems is to select, optimize, and measure one or more metrics throughout the development pipeline. Common metrics, where applicable, include accuracy, precision, recall, ROC curves, confusion matrices, mean squared error (MSE), and/or mean absolute error (MAE) (Handelman et al., 2019). While the optimization of metrics in software development is not unique to AI, AI is exceptionally good at performing optimizations. Although properly defined and comprehensive metrics can yield impressive quantitative results, excessive optimization of inadequate metrics can result in manipulation, gaming, and/or a focus on short-term quantities, in addition to other potential negative consequences (Thomas & Uminsky, 2020). Utilizing incomplete and/or misleading metrics to test and evaluate AI systems is therefore fraught with risk. Additionally, the phases of this pipeline are often viewed as distinct stages and as a result teams across the phases of development may be siloed from one another. This can lead to barriers in interpreting and responding to information derived from T&E metrics.

Today, questions around the design, development, implementation, and sustained management of AI are examined across a variety of fields including software engineering, human-centered design, computer science, and systems engineering. We believe that to build AI as well as it can be done, a whole-systems approach is needed. As stated by Ackoff and Wardman (2016), "when you take a system apart, it loses all of its essential properties." In this paper, we outline holistic considerations for testing and evaluation and aim to extend beyond common practice by capturing cross-disciplinary perspectives on AI engineering, acknowledging the volatility of relying heavily on metrics, and addressing the unique challenges of working with numerous evolving, interconnected system components.

## Characteristics of AI Systems

AI Engineering is a field of research and practice that combines cross-disciplinary perspectives to create AI systems in accordance with human needs for stakeholder outcomes (CMU SEI, 2022). When thinking about measurement of systems, it's important to start with defining the attributes of a system that are desired. Here, we draw on the three pillars of AI Engineering to guide our thinking on test and evaluation strategy:

1. Human-Centered AI: Implementing AI in context requires a deep understanding of the individuals who intend to use and interact with the system. From a human-centered perspective, systems should be evaluated to assess the alignment with humans, their behaviors, and values, as well as the utility of systems to achieve stakeholder-driven outcomes.
2. Scalable AI: Many current AI and ML methods are compute-intensive, expensive, and time-consuming to develop, necessitating consideration of how to scale AI and ML techniques to real-world size and complexity. In the context of scalability, evaluation lenses could include how AI infrastructure, data, and models may be reused across problem domains and deployments and increase performance to support operational needs. Another evaluation lens could include how effectively a system scales up to support more needs at the enterprise level or scales down to enable capabilities in edge contexts.
3. Robust and Secure AI: AI and ML introduce new system failure modes, vulnerabilities, and attack vectors and change over time. Assurance is needed that systems will work as expected when faced with uncertainty or threat. The Robust and Secure AI pillar provides the context for how we develop and test AI systems to ensure resilience across contexts and when encountering new phenomena over time.

With these pillars in mind, we examine core considerations that AI engineers and AI engineering teams must take into account when developing test and evaluation strategies for AI systems. Research driven by these considerations is called for to push the state of the art forward and build a discipline around AI engineering. Due to the prevalence of ML systems in use today, we will provide many of the considerations in the context of ML.

## Key Questions when Assessing AI Systems

To steer discussion, research, and implementation of T&E strategies for AI systems, we propose the following six key considerations. These questions capture cross-disciplinary AI engineering concerns related to deriving actionable lessons from testing, identifying sources of risk and uncertainty, and evaluating system suitability for different use-cases and end-users.

### 1. What are you intending to test (and learn)?

While many of the questions listed below are not unique to AI systems in the abstract, each becomes unique within the context of AI systems. Unlike in traditional software, the answers to each of these questions are typically less definitive due to the non-deterministic nature of AI. An AI system may perform as expected given one input but behave in an unintended manner on a similar input; testing cannot be exhaustive due to the singular nature of any one data point. As a result, there is a limit to the possible coverage of testing in the context of AI systems.

A helpful starting point for AI engineers is to consider what possible sources of uncertainty exist in a system of interest. In image classification, for instance, the system makes class predictions with varying degrees of confidence, but typically only the prediction with the highest confidence is returned to the end-user. Additional information about uncertainty is helpful during the testing process. Some classes may be hard to distinguish between, even for humans, in which case the model's limitations mirror human performance. Other mix-ups, for instances between classes that look completely dissimilar, could be indicative of more critical model deficiencies.

Brainstorming potential sources of risk can be another informative practice. What are the potential negative and positive outcomes that the system can produce? How can you evaluate for these outcomes, both indirectly and directly? T&E allows for a thorough exploration of the model's behavior before it is deployed and is an opportunity to measure and mitigate sources of risk. This practice is important because it upholds the principles of ethical AI, but it also may be a requirement if a governing entity requires regulatory or legal tests.

Overall, as AI engineers start their exploration into assessment of systems, there are many directions in which they could go and realistically, they have both limited time and resources. Setting clear intentions and learning goals around T&E can help teams prioritize what they are investing resources into at different times and ensure they are extracting actionable information from the testing that is performed.

## 2. What logistical challenges might you encounter during testing?

Developing a well-documented plan for handling the logistical challenges of T&E for AI systems is another task that is necessary to perform early on. Evaluations can be performed at a variety of levels. AI systems require traditional software T&E-style verification that code is performing properly and is free of bugs; this form of testing can be performed by standard software developers. However, aspects of AI systems that require expertise, such as interpreting the meaning of increased uncertainty in deployment, recognizing the emergence of new classes, or determining efficacy in human-machine teaming contexts may require specialized team members to become involved in the evaluation. Outlining all relevant evaluation concerns and assembling a diverse team to tackle T&E tasks spanning different risk levels and content areas is a crucial step in producing a robust T&E strategy.

Due to the cyclical, interconnected nature of ML lifecycle phases, issues discovered through T&E can have wide-reaching effects for ML systems and plans to mitigate these impacts will likely require communication across phases. The development of ML systems has been mapped to various lifecycle frameworks. Andrew Ng, for instance, educates practitioners to follow a four-part ML lifecycle that includes: (1) Scoping the project, (2) Collecting data (3) Training a model, and (4) Deploying in production (DeepLearningAI, 2021). Garcia et al. (2018) contends that the ML lifecycle consists of context, defined as "all the information surrounding the use of data in an organization," plus three phases: (1) Pipeline Development, (2) Training, and (3) Inference. While differences in proposed frameworks exist, what is true across frameworks is that the ML lifecycle consists of numerous phases intertwined via feedback loops.

Although testing is often represented as either a distinct phase of the AI system lifecycle or as a component of the "training phase", we advocate that T&E is most effective when integrated throughout all phases. When implementing AI systems, teams of humans must engage in continuous oversight and frequently reflect on the questions: What are we doing? Why are we doing it, and for whom? (Barmer et al., 2021). Accounting for T&E considerations across every stage of AI system development and deployment supports the rapid, iterative development of robust, ethical mission capabilities (JAIC, 2020). Selecting metrics that accurately assess the system's ability to fulfill mission goals and using these metrics throughout training, for instance, will guide the system towards meeting stakeholder needs.

A common misconception is that testing is overly time-consuming, while the process of fixing errors, especially late in system development, is what actually absorbs time (Kohavi et al., 2009). Developing a culture of frequent testing and a practice of addressing errors as they arise is an effective method for catching issues early on and preventing a build-up of problems. As stated by Thomke (2020), "Culture—not tools and technology— prevents companies from conducting the hundreds, even thousands, of tests they should be doing annually and then applying the results."

In settings where personnel across the pipeline work together closely, building T&E into all stages of development is realizable. However, when different stakeholders are siloed across the pipeline, such as designers, data scientists, software engineers, machine learning researchers, and operations teams, challenges communicating between roles can be a source of ML "mismatch" (Lewis et al., 2021). Facilitating effective communication across roles on AI teams and throughout organizations to ensure that errors and problems found during T&E are addressed properly is a challenge, especially when elements of the ML pipeline may be handled by different organizations or teams. Methods for addressing errors in T&E may include system rollback (especially in high-risk, mission-critical contexts) or revising the training dataset to account for new information about model performance (Dunnmon et al., 2021).

Constraints on access to information across development and deployment pipeline stages should also be accounted for. For example, if provenance or other details about the training data are not accessible by the T&E team this can make it difficult to detect bias and uncertainty or to know what vulnerabilities or edge cases to evaluate. Determining and documenting what aspects of the model and pipeline are within the scope of testing, and in turn what issues and topics can realistically be addressed, is an important step in fleshing out a T&E strategy. Limitations on T&E personnel's access to data or training/deployment specifics can be significant obstacles that may require modifications to team structures and/or documentation procedures.

## 3. What are your biggest sources of risk?

In the context of AI-enabled systems, it's important to frame risk, or the possibility of suffering loss, in the context of the role that AI is performing within the system (Dzombak et al., 2021; Dorofee et al., 1996). For instance, if an AI component stops working or begins to operate poorly, how will this impact the system's overall ability to perform its task?

What does poor operation look like and how can it be measured? Enumerating potential threats to the system, the likelihood of each threat occurring, and the impact of each of these threats early on will provide the T&E team with an estimate of risk that can guide the focus of testing (Alberts & Dorofee, 2010).

Traditional methods for estimating the impact of loss (Kambic et al., 2020; Tucker, 2020) can be applied to AI systems, but estimating the likelihood of loss in the context of AI is an open challenge. Nascent methods for determining the quantitative likelihood of loss can be pulled from an emerging body of work related to AI threat modeling and vulnerabilities (Biggio & Roli, 2018; Beieler, 2019; Householder et al., 2020a; Householder et al., 2020b; MITRE, 2020). In the absence of quantitative estimates, qualitative assessments from domain experts can be leveraged to gauge the relative importance of threats.

Since AI systems have the potential to be used for different tasks, understanding the specific use-cases for which the system will be employed can sharpen the objectives of risk-related testing. For example, consider an overhead object detection system designed to identify vehicles of interest to military personnel. Two potential use-cases for the system could include (1) reconnaissance, a mission that is limited in time and scope, or (2) surveillance, a longer-term mission with less time pressure. In the reconnaissance use-case, it's important that the false positive rate is kept low because spurious hits could overwhelm the end-user as they make time-critical decisions. In the surveillance scenario, on the other hand, the rate of false negatives is a bigger concern because time constraints are weaker and unflagged incidents could be costly. Considering the specific use-cases for a system and the risks involved in these scenarios can help determine which metrics are most relevant.

### 4. What is the meaning behind your metrics?

A challenge in interpreting and selecting metrics for T&E is determining their meaning and impact in *context*. Donella Meadows (1998) stated: "Indicators arise from values (we measure what we care about) and they create values (we care about what we measure)". What is the overall value provided by the AI system and what kinds of measurements can be used to assess progress towards providing this intended value? What impact will prioritizing certain metrics have on the system development? Often, teams implement the measurement systems that they have knowledge of, whether or not they provide the needed meaning. In AI systems, garnering meaning from metrics is complicated by factors such as system complexity, risk, and audience.

AI systems have exceptionally powerful optimization capabilities, therefore the optimization of metrics that do not align with intended values can build systems with impressive test results but produce behaviors that are both unintentional and consequential. For example, Facebook used metrics such as time spent on the platform and the number of posts that users interacted with as proxies for measuring progress towards their goal of facilitating social connection. In the process of optimizing these metrics, their algorithm learned to show users posts that upset or anger them. While the engagement metrics may have improved, their progress towards the goal ultimately suffered. This example demonstrates how defining system goals and choosing metrics for T&E that truly support these values is critical.

While intended system value may be clear, goals towards achieving this value, as well as the metrics for measuring progress towards goals, can be competing or misaligned. For instance, a facial recognition system will likely have to make tradeoffs between achieving high efficiency and fairness across protected groups. Tradeoffs may involve other ethical issues such as privacy, transparency, and accountability (Amarasinghe et al., 2021). Identifying and assessing tradeoffs between metrics is a challenge which remains ongoing throughout the ML system lifecycle.

Assurance that metrics are calculated accurately is a prerequisite to deriving meaning from measurements. Standard metrics such as accuracy or false positive rates are relatively easy to verify, especially in the presence of clear ground-truth labels. Confidence scores, on the other hand, are a more complex and often unverified metric. While confidence scores can be a valuable source of information about model performance during T&E, these estimates are only useful if they have been calibrated to suggest the true correctness likelihood (Guo et al., 2017). Mechanisms for producing front-facing metrics, such as confidence scores, must be validated prior to deployment to ensure that end-users are given precise information when interacting with the system.

When interpreting performance metrics for non-technical audiences specifically, another set of challenges and opportunities arise. Scores such as F1 or AUC ROC can be difficult to interpret without a technical background in AI; how can these metrics be translated into plain English in the context of the problem at hand? Meaning must be derivable from system metrics not only by ML practitioners, but by other key collaborators involved in designing and reviewing the system. Efforts must also be made to avoid information overload by condensing relevant information and presenting results in a clear, simple, and balanced manner that is accessible for non-technical stakeholders (IDF, 2020).

Additionally, it's important to consider what metrics requirements really mean and how they align with project objectives. For instance, if a decision threshold was used, how can this cutoff be justified and was this decision appropriate for the goal? Fan and Lin (2007) discuss how performance metrics can be improved by changing decision thresholds. Modifying thresholds, however, can result in a selection rate that does not make sense for your problem. Perspectives from domain experts, where applicable, can help shape discussions around metric expectations and parameters including threshold.

## 5. How are you dealing with the scale and level of complexity of your system?

A significant challenge facing AI system developers today is how to create systems that can operate across a variety of domains and use cases. Success is hard enough to achieve when operating AI systems in closely controlled development and laboratory environments, and even more challenging when considering scale and system complexity.

Achieving the development and deployment of robust and secure AI systems requires the creation of new T&E strategies that take scope into account. To thoroughly evaluate system performance, T&E teams must acknowledge that an AI system will not exist in a vacuum; consider how the system operates and what kinds of interactions will take place between the system and sub-systems or contextual systems of interest. What inputs will the system receive and what outputs are expected? What impact will faulty results or predictions have on downstream components? How does the system respond to invalid inputs? This context can inform expectations about system behavior and sources of potential risk and, in turn, guide the selection of appropriate metrics and methods for addressing these requirements.

Furthermore, it's important to consider and routinely reevaluate the setting(s) in which the AI system will be employed and what use in these environments entails. What differences are there between the local test environment versus global implementation contexts? In deployment, a system may require different compute resources to meet increased demand or operate on a delayed retraining schedule. Likewise, the system may encounter different real-world relationships between inputs and outputs, unexpected input data, or distinct types of end-users. To prepare for diverse use-cases, training and testing data must address a variety of real-world scenarios.

That being said, comprehensive coverage is likely unattainable. For complex systems, it's impossible to generate a complete list of scenarios in which the system may fail (Doshi-Velez & Kim, 2017). Instead of working towards a "perfect" system, T&E teams can build confidence in an AI system through rigorous testing, evaluation, verification and validation (TEV&V) incorporated throughout the system's lifecycle. While in traditional software operational metrics are the primary concern for system monitoring, in the realm of AI engineering performance metrics are also essential (Huyen, 2022). Monitoring a system in deployment contexts and tracking new behavior such as data drift will provide teams with the information needed to retrain a system to meet emerging needs or to tune performance expectations to reflect new requirements.

## 6. How are you evaluating for bias and other unintended behaviors?

To ensure that a system is responsible and equitable, it must be vetted during T&E for unintended and/or negative consequences on the humans who will be impacted by the system. Across all phases of AI system development, it's important to keep in mind who will be engaging with the system, directly or indirectly, and what they will potentially gain or lose through their interactions. If the system is designed to support or replace an existing process, tests that reflect existing expectations and best practices in deployment should be developed alongside domain experts. What pain-points for the user within the current system, and to what degree will the new system improve upon or change how such issues manifest? Likewise, what potentially negative tradeoffs exist in the new system and how will they impact the user?

ML models learn the relationships explicitly or implicitly embedded within their (potentially historical) training dataset; as a result, they have the potential to pick up on undesirable correlations between inputs and outputs. The existence of unintended relationships between features in the training dataset can cause the model to "learn" the wrong thing altogether. Geirhos et al. (2021) describes one such example in which an image classifier identifies cows based on the presence of grass in the image; while the model achieved high performance accuracy, the model failed on examples in which a cow was pictured in a new setting. When working with human-related inputs, such as in facial recognition systems, correlations can include racial or gender bias, among other demographic disparities. Unintentional correlations in training data can have real-world impacts; New Jersey's pretrial risk assessment algorithm, for instance, was trained on data that "reflects racial and ethnic disparities in policing, charging, and judicial decisions" and, as a result, made decisions that "perpetuate racial inequalities" with regards to detainment (Simonite, 2020).

Analyzing training data directly is a powerful method for detecting and preventing biases and other unwanted correlations. The Gender Shades project provides an illustrative example of how accuracy on a per-demographic basis can vary widely and comparing model performance across different data slices can reveal underlying system biases (Buolamwini & Gebru, 2018). The detection and mitigation of bias is a crucial concern for AI systems working with human subjects, but systems designed for other input types can also exhibit biases, such as producing better results for certain languages or geographic inputs than others. Careful analysis of training, evaluation, and testing data in the early stages of model development is a crucial preventative measure towards smart testing and evaluation. Repeated analysis across stages, especially if data is collected in deployment and used to retrain the model, is another important quality check. Explainability techniques can also be used to periodically probe the model and determine which features are the most important factors in determining system outputs. Auditing of a system before its adoption is critical to prevent unwanted consequences.

An important caveat to consider throughout bias mitigation efforts is the risk of accidentally masking unfair behavior through selected metrics. The Propublic Machine Bias study (Angwin et al., 2016) provides a case study of how sparse metrics can conceal discriminatory model behavior. In the study, researchers examined a "fair" model for predicting recidivism based on criminal history that received similar accuracies across different racial groups. Upon closer examination, however, researchers discovered that Black defendants were twice as likely to be falsely identified as recidivists than white defendants. The decision to use accuracy alone to identify bias resulted in real-world racially discriminatory practices; this study illustrates the importance of conducting in-depth analysis of model behavior, including an investigation of how the model fails.

## Iterating on a T&E Strategy

Robust T&E hinges on the realignment of priorities in response to emerging needs. After a first attempt at evaluation, it's important to reconsider what the main tradeoffs are within the evaluation strategy and determine if there are any important system or model attributes that are currently unaccounted for. Throughout the lifecycle of the model or system, different needs may appear as data inputs and/or expected outputs change. Proper monitoring of the system (often considered the final stage of the ML system pipeline) is necessary to recognize shifts such as data drift or concept drift. Information gleaned about current project needs through monitoring can be used to ensure that the T&E procedure covers relevant risks and concerns.

For instance, imagine an ML speech-to-text model for making song requests as part of an oldies music streaming system. Developers focus on training and testing a model that achieves high accuracy on speech samples from users who are middle-aged and older, as this is the platform's target demographic. The system performs well for the first six months but sees an increase in teenage users in the second half of the year and a corresponding drop in average accuracy. While the initial goal for T&E was to ensure that the system achieved high accuracy for their target age group, the top priority will now likely shift towards achieving acceptable accuracy across all age groups. Dataset curation for both training and testing must grow to include samples from teenage speakers and, in anticipation for future new users, developers should consider including a larger range of speech samples across different demographics (e.g., age, dialect).

Other possible changing project needs could include increased or decreased scale, the emergence of adversaries, or the introduction of a new class. T&E considerations will likely fluctuate in priority depending on these needs. Keeping inventory of both the aspects of the system that have been thoroughly tested as well as open topics for future testing will be crucial to developing and maintaining a robust and up-to-date strategy.

## Conclusion

To conclude, testing for accuracy alone is not enough to assess the correctness or quality of a ML model. To engineer robust and secure, scalable, and human-centered AI systems T&E needs to account for potential sources of risk and uncertainty early on and incorporate testing measures that address these concerns across all stages of development and deployment. This approach differs from typical AI T&E approaches that view testing as a stage within a linear pipeline and instead opts for a holistic vision of testing that considers the connections between phases of the model lifecycle. Since comprehensive testing is impossible for AI systems, it's important to determine the intentions behind testing and to make informed tradeoffs. Maintaining documentation of process, iterating on T&E strategies in response to emerging requirements, and developing diverse teams to handle varied testing responsibilities are practices that can improve both the depth and breadth of testing. As AI engineering best practices continue to evolve, the delta between traditional systems and AI systems will be further explored and addressed.

## Acknowledgements

The authors would like to acknowledge Carol Smith for her invaluable input framing human-centered concerns for consideration six.

This material is based upon work funded and supported by the Department of Defense under Contract No. FA8702-15-D-0002 with Carnegie Mellon University for the operation of the Software Engineering Institute, a federally funded research and development center.

# Copyright